\documentclass[superscriptaddress,dvipsnames,nofootinbib,amsmath,amssymb,prd,notitlepage,prd,showpacs,twocolumn]{revtex4-1}
\usepackage{color}
\usepackage{bm}
\usepackage{graphicx}
\usepackage{amsmath}
\usepackage{amssymb}
\usepackage{enumitem}
\usepackage{hyperref}
\usepackage{subfigure}
\usepackage{array}
\usepackage[english]{babel}
\usepackage{tensor}
\usepackage{comment}
\usepackage[normalem]{ulem}
\usepackage[dvipsnames]{xcolor}

\usepackage{fontawesome}

\def\be{\begin{equation}}
\def\ee{\end{equation}}
\def\ba{\begin{eqnarray}}
\def\ea{\end{eqnarray}}
\def\l{\left}
\def\r{\right}
\def\f{\frac}

\allowdisplaybreaks

\def\nn{\nonumber}

\begin{document}

\title{N-body Simulations of Large-Scale Structure in \\ the Generalized Cubic Covariant Galileon Model}

\author{Lu\'is Atayde}
\email{luisbbatayde@gmail.com}
\affiliation{
 Instituto de Astrofis\'ica e Ci\^{e}ncias do Espa\c{c}o, Faculdade de Ci\^{e}ncias da Universidade de Lisboa, Edificio C8, Campo Grande, P-1749016, Lisboa, Portugal }
 \author{Noemi Frusciante}
 \email{noemi.frusciante@unina.it}
 \affiliation{Dipartimento di Fisica ``E. Pancini", Universit\`a degli Studi  di Napoli  ``Federico II", Compl. Univ. di Monte S. Angelo, Edificio G, Via Cinthia, I-80126, Napoli, Italy}
 \affiliation{INFN Sezione di Napoli, Università degli Studi di Napoli “Federico II”,\\Compl. Univ. di Monte S. Angelo, Edificio G, Via Cinthia, I-80126, Napoli, Italy}
\author{Baojiu Li}
\email{baojiu.li@durham.ac.uk}
\affiliation{
 Institute for Computational Cosmology, Department of Physics, Durham University, South Road, Durham, DH1 3LE, UK}
 
\begin{abstract}
 We present the first N-body simulations of structure formation in the
Generalized Cubic Covariant Galileon (GCCG) model. This theory extends the
cubic covariant Galileon through power-law kinetic and cubic derivative
interactions and admits tracker solutions leading to late-time cosmic
acceleration. Previous studies of GCCG have mostly focused on the background, linear perturbations, or semi-analytic nonlinear prescriptions. Here we implement the nonlinear scalar-field equation in the \texttt{ECOSMOG} adaptive-mesh refinement code, allowing us to follow the coupled evolution of matter
clustering and Vainshtein screening in the fully nonlinear regime.
 We quantify the impact of GCCG on the nonlinear matter power spectrum and
compare the simulation results with predictions from the halo-model reaction
approach. For the parameter choices considered, the GCCG model enhances the
matter power spectrum relative to the corresponding QCDM cosmology, with the
effect increasing towards low redshift and reaching approximately $7\%$ at
$z=0$ in the transition to the nonlinear regime. At smaller scales, the
enhancement decreases as a consequence of Vainshtein screening. We find that
the reaction framework captures the qualitative behaviour of the simulations, while residual differences appear on deeply nonlinear scales. We also analyse
the abundance of dark matter haloes, finding an enhancement relative to QCDM
that becomes more pronounced towards lower redshift and in the high-mass tail.
These simulations provide the
first nonlinear calibration of structure formation in GCCG and establish the
range of validity of efficient semi-analytical predictions for applications
to forthcoming large-scale-structure surveys.
\end{abstract}

\date{\today}
\maketitle

\section{Introduction}

The late-time accelerated expansion of the Universe, inferred from observations of Type Ia supernovae~\cite{SupernovaSearchTeam:1998fmf, SupernovaCosmologyProject:1998vns} and reinforced by cosmic microwave background (CMB) measurements and baryon acoustic oscillations (BAO)~\cite{Planck:2018vyg, SDSS:2005xqv, BOSS:2016wmc}, remains one of the most important challenges in modern cosmology. The standard $\Lambda$CDM model, anchored by a cosmological constant ($\Lambda$), is remarkably consistent with current observations, despite a slight tension in the measurement of the Hubble parameter \cite{DiValentino:2020zio}. It also lacks a compelling physical origin and faces severe fine-tuning and coincidence problems~\cite{Weinberg:1988cp,Carroll:2000fy}. These theoretical challenges have invigorated searches for alternative explanations, including modifications of General Relativity (GR) and dynamically evolving dark energy (DE) models \cite{CANTATA:2021asi}.

Recent observational advances by the Dark Energy Spectroscopic Instrument (DESI) have further motivated this effort. Analyses of DESI DR1 and DR2, incorporating over 14 million galaxies and quasars, have been combined with external datasets including Type Ia supernovae, weak lensing, and the CMB. These joint analyses provide growing evidence for a time-dependent DE equation of state, with deviations from a cosmological constant detected at the $2.8$--$4.2\sigma$ level~\cite{DESI:2024uvr,DESI:2024lzq,DESI:2025wyn,DESI:2025zgx,DES:2026jmi}. Bayesian model comparison consistently favours dynamical DE models, such as quintessence and phantom-like scenarios, over $\Lambda$CDM, suggesting that cosmic acceleration may not be driven by a true constant, but instead by an evolving scalar field \cite{Ye:2024ywg,Wolf:2025jed,Wolf:2024stt}. 

In this context, scalar-tensor theories, and in particular the Galileon (or equivalently Horndeski) class~\cite{Horndeski:1974wa, Deffayet:2011gz,Kobayashi:2011nu}, provide a well-motivated framework for constructing cosmological models with dynamical DE and modify the gravitational interaction at large scales. These are four-dimensional theories with a single scalar field and second-order equations of motion, thus avoiding Ostrogradsky instabilities \cite{Ostrogradsky:1850fid,Woodard:2015zca}. A notable subset is the cubic shift symmetric Galileon, in which the scalar field $\phi$ enters the action through a kinetic term, $G_2(X)$ and a cubic interaction $G_3(X)\Box\phi$, with $X \equiv (\nabla\phi)^2$. These terms maintain the second-order nature of the field equations in curved spacetime and are consistent with the gravitational wave speed constraint from GW170817~\cite{LIGOScientific:2017vwq}.

A specific model is the Generalized Cubic Covariant Galileon model \cite{DeFelice:2011bh}, which exhibits a rich phenomenology. With power-law choices for $G_2(X)$ and $G_3(X)$, the model admits \textit{tracker solutions} that dynamically drive late-time acceleration without requiring fine-tuned initial conditions. At the background level, the effective DE equation of state can satisfy $w_{\rm DE} \lesssim -1$, providing a potential resolution to the Hubble tension by increasing the inferred value of $H_0$ from CMB data alone~\cite{Frusciante:2019puu}. At the perturbation level, the model introduces scale-dependent modifications to the growth of structure, offering observational signatures in weak lensing, redshift-space distortions, integrated Sachs-Wolfe CMB and galaxy clustering~\cite{Frusciante:2019puu,Giacomello:2018jfi,Kable:2021yws}. Comprehensive analyses combining CMB, BAO, SNe Ia, and local $H_0$ measurements show that GCCG performs competitively with $\Lambda$CDM and can even produce a nonzero lower bound on the sum of neutrino masses, $\sum m_\nu$~\cite{Frusciante:2019puu}. 

However, most investigations of the GCCG model have been restricted to background and linear perturbation regimes. To extract meaningful predictions for upcoming high-precision cosmological surveys such as \textit{Euclid}, it is essential to understand the behaviour of the theory in the non-linear regime, where screening mechanisms such as the Vainshtein effect \cite{Vainshtein:1972sx,Joyce:2014kja} operate to suppress deviations from GR in high-density regions. Accurately modelling these effects requires solving the full scalar field dynamics in a non-linear cosmological context.

Recent progress has been made in extending semi-analytic techniques to Galileon-type models. In particular, Ref. \cite{Atayde:2024tnr} applied the halo model reaction framework \cite{Cataneo:2018cic} to the GCCG model, adapting the \texttt{ReACT} code \cite{Bose:2020wch,Bose:2021mkz} to account for its additional K-essence and cubic derivative power-law terms. The method was validated \cite{Atayde:2024tnr} against full $N$-body simulations \cite{Barreira:2013eea} in the limit in which the GCCG model resembles the original Cubic Covariant Galileon model \cite{Deffayet:2009wt}, achieving $\sim$5\% accuracy across a wide range of scales and redshifts, and used to forecast percent-level constraints on the extra GCCG parameters for forthcoming spectroscopic and photometric surveys, such as \textit{Euclid} \footnote{www.euclid-ec.org} and SKAO \footnote{Square Kilometer Array Observatory: https://www.skao.int} \cite{Atayde:2024tnr}. These results highlight the potential of reaction-based approaches as efficient surrogates for large simulation suites, provided that they are carefully calibrated against high-fidelity numerical results \cite{Bose:2024qbw}.

To this end, $N$-body simulations are indispensable. These simulations self-consistently evolve the matter distribution and scalar field, allowing one to capture the feedback between gravitational collapse and modified gravity (MG) effects. Several simulation frameworks, such as \texttt{ECOSMOG}~\cite{Li:2011vk}, \texttt{MG-GADGET}~\cite{Puchwein:2013lza}, and \texttt{ISIS}~\cite{Llinares:2013jza}, have been developed to study various modified gravity models, including $f(R)$~\cite{Li:2011vk}, DGP \cite{Schmidt:2009sv,Schmidt:2009sg}, Galileon-type theories \cite{Li:2013nua,Barreira:2013eea} and even model independent approaches such as the effective field theory of DE \cite{Hassani:2020rxd,Nouri-Zonoz:2025cul,Ganjoo:2026ugf,Woodcock:2026lpv}. 

In this work, we take a key step towards probing the non-linear phenomenology of the GCCG model by implementing the model within the \texttt{ECOSMOG} $N$-body simulation code. This marks the first such implementation of GCCG and enables a fully nonlinear treatment of the scalar field evolution and Vainshtein screening during cosmic structure formation. Our aim is to quantify the impact of GCCG on dark matter halo formation, the non-linear matter power spectrum, and other large-scale structure observables. These predictions will provide the necessary theoretical input for interpreting data from future high-precision surveys such as \textit{Euclid}, the \textit{Vera C. Rubin Observatory} (LSST) \footnote{https://www.lsst.org}, and the \textit{Nancy Grace Roman Space Telescope} \footnote{https://roman.gsfc.nasa.gov}. Ultimately, this work will lay the groundwork for testing GCCG in the deeply non-linear regime.

The paper is organised as follows. In Section \ref{sec:model}, we review the GCCG model, discussing in detail the background evolution in Section \ref{sec:tracker} and the nonlinear dynamics implemented in the N-body code in Section \ref{sec:nonlinear}.  Section \ref{sec:Nbody} presents the setup of the numerical simulations, while Section \ref{sec:results} is devoted to the presentation and discussion of the results. Finally, we summarise our conclusions in Section \ref{sec:conclusion}.

\section{The model}\label{sec:model}

We consider the shift symmetric cubic Galileon theory described by the action~\cite{Deffayet:2010qz,Kobayashi:2010cm}
\ba\label{action1}
S=\int d^4x\sqrt{-g}\l(\f{M_{\rm pl}^2}{2} R +\mathcal{L}_\phi+ \mathcal{L}_m(g_{\mu\nu},\chi_i)\r), \nn \\  
\ea
where the Lagrangian which includes modifications to the gravity sector is
\be
\mathcal{L}_\phi=G_2(X)+G_3(X)  \Box\phi\,,
\ee
with $M_{\rm pl}^2$ being the Planck mass, $R$ is the Ricci scalar, $g$ is the determinant of the metric $g_{\mu\nu}$, 
$G_i$ are free functions of  $X=\partial_\mu \phi\partial^\mu\phi$ and $\phi$ is the scalar field. $\mathcal{L}_m$ stands for the matter Lagrangian for all matter fields, $\chi_i$.

The variation of the action with respect to the metric gives the following field equations \cite{Kobayashi:2011nu}
\begin{align}
    &G_{2X} \nabla_{\mu}\phi \nabla_{\nu}\phi - \frac{1}{2} G_2 g_{\mu \nu} + G_{3X} \Box \phi \nabla_{\mu} \phi \nabla_{\nu} \phi \nonumber \\ 
    &- \nabla_{( \mu} G_3 \nabla_{\nu )} \phi + \frac{1}{2} g_{\mu \nu} \nabla_{\lambda} G_3 \nabla^{\lambda} \phi + \frac{M^2_{\rm pl}}{2} G_{\mu \nu} = T_{\mu\nu}\,, \label{eq:eq1}
\end{align}
where $G_{iX} \equiv \partial G_i / \partial X$, $G_{\mu \nu}$ is the Einstein tensor and $T_{\mu\nu}$ is the stress energy tensor. The variation with respect to the scalar field gives:
\be
  \nabla^{\mu} \left[ 2 G_{2X} \nabla_{\mu} \phi + 2 G_{3X} \Box \phi \nabla_{\mu} \phi - G_{3X} \nabla_{\mu} X \right] = 0\,, \label{eq:eq2}
\ee

For the matter sector holds the conservation of the Stress Energy Tensor:
\begin{equation} \label{eq:continuityequation}
 \nabla^\mu T_{\mu\nu}=0\,.
\end{equation}
We consider barotropic fluids with pressure $p_m$ and density $\rho_m$ related by $p_m=w\rho_m$, with $w=0$ for baryons and cold dark matter.

We specify the forms of the $G_i$ functions to be the ones of the Generalized Cubic Covariant Galileon (hereafter GCCG)~\cite{DeFelice:2011bh}:
\ba
&&G_2=-c_2  \alpha _2^{4 \left(1-p_2\right)} \l(-X\r)^{p_2},\quad G_3=-c_3  \alpha _3^{1-4 p_3}  \l(-X\r)^{p_3}\,,\nn\\
&&
\ea
 where $c_i$ and $p_i$ are  dimensionless constants and $\alpha_i$ are constants with dimensions of mass, defined as
\be
\alpha_2=\sqrt{H_0 M_{\rm pl}}\,,\quad \alpha_3=\left(\frac{M_{\rm pl}^{1-2 p_3}}{H_0^{2 p_3}}\right){}^{\frac{1}{1-4 p_3}}\,,
\ee
with $H_0$ being the Hubble constant.  We set $c_2=1/2$ without loss of generality~\cite{Barreira:2013xea,2014JCAP...08..059B,Renk:2017rzu}. 

In the following, we will review the background and non-linear perturbation evolutions. We will consider a perturbed  Friedmann-Lemaître-Robertson-Walker (FLRW) metric in Newtonian gauge:
\be
\label{metric}
ds^2 = -\left(1 + 2\Psi\right)dt^2 + a(t)^2\left(1 - 2\Phi\right)\gamma_{ij}dx^idx^j,
\ee
where $a(t)$ is the scale factor, $\Psi(t,x_i)$ and $\Phi(t,x_i)$ are the gravitational potentials representing perturbations of the metric components and are functions of cosmic time, $t$ and space, $x_i$.

\subsection{Background evolution: the tracker solution}
\label{sec:tracker}

At the background level, the GCCG model shows a tracker solution, which we will use to solve the equations and evolve the main background quantities. It is given by~\cite{DeFelice:2011bh}
\be\label{tracker}
H^{2q+1}\psi^{2q}=\zeta H_0^{2q+1}\,, 
\ee
where $H\equiv \dot{a}/a$ is the Hubble function and dots stand for derivatives with respect to cosmic time, $t$, and $H_0=H(a=1)$ is its present time value,
$\zeta$ is a dimensionless constant. We introduced a dimensionless constant $q\equiv(p_3-p_2) +1/2$ 
and a dimensionless scalar field $\psi$  defined as
\be
\psi\equiv\f{1}{M_{\rm pl}}\f{d \phi}{d\ln a}\,.
\ee

The Friedmann equations can be obtained from the equation \eqref{eq:eq1} and  with the above tracker solution  they read \cite{Frusciante:2019puu}: 
\be\label{eq:background}
\l(\f{H}{H_0}\r)^{2+s}=\Omega_\phi^0+\f{\Omega_m^0}{a^3}\l(\f{H}{H_0}\r)^{s}\,,
\ee
\be
\frac{\dot{H}}{H^2}=-\frac{\l(\frac{H_0}{H}\r)^2 \Omega_m^0 a^{-3}(-3-s)+(s+2)}{\l(\frac{H_0}{H}\r)^2s\Omega_m^0 a^{-3}-(s+2)}-1\,.
\ee
We defined: $s=p_2/q$; $\Omega_{m}^0\equiv \rho_{m}^0/3M_{\rm pl}^2H_0^2$, which is the density parameter at present time for matter component (m$=$cold dark matter$+$ baryonic matter) and the density of the scalar field at present time, $\Omega_\phi^0$  which is:
\ba\label{flatness}
\Omega_\phi^0=1-\Omega_{\rm m}^0=c_3 (2 s\,q+2 q-1) \zeta ^{s+1}-\frac{1}{6} (2 s\,q -1)\zeta ^{s}\,.\nn\\
\ea
The latter is obtained by evaluating eq. \eqref{eq:background} at present time ($a=1$). We have also used the solution for the matter behaviour $\rho_m \propto a^{-3}$ which is obtained by solving the continuity equation, $\dot{\rho}_m+3H\rho_m=0$, from eq. \eqref{eq:continuityequation}.

It is possible to show that:
\ba
\zeta=\l(6\Omega_\phi^0\r)^{\f{1}{s}}\,,\quad c_3=\f{1}{3}\f{s \, q }{\l(6\Omega_\phi^0\r)^{\f{1}{s}}(2s \, q+2q-1)}\,\,.
\ea
These relations demonstrate that the GCCG model has only two extra free parameters, i.e. $\{s, q\}$, with respect to $\Lambda$CDM.  
 In the limit $s=2$ and $q=1/2$ the model reduces to the Cubic Galileon \cite{Deffayet:2009wt}. The parameters $\{q,s\}$ are restricted to be positive due to stability conditions, i.e. ghost and gradient requirements \cite{DeFelice:2011bh,Frusciante:2019puu}.

\subsection{Non-linear evolution}
\label{sec:nonlinear}

In this section we review the non-linear equations for the GCCG. We adopt the quasi static approximation (QSA) which assumes that time derivatives of the perturbed quantities can be neglected
compared with their spatial derivatives. Under this approximation  the non-linear equation for the Poisson equation reads \cite{Albuquerque:2024hwv}:

\begin{equation}
    \label{eq:poisson}
\nabla^2\Psi=
4\pi G a^2\rho_{\rm m}\delta_{\rm m}
+
\alpha_{\rm B}\frac{H}{\dot{\phi}}\nabla^2\delta\phi,
\end{equation}
where
$
\delta_{\rm m}\equiv\frac{\delta\rho_{\rm m}}{\rho_{\rm m}}
$
is the matter density contrast and $\delta\phi$ denotes the scalar-field perturbation. The function $\alpha_{\rm B}$, commonly referred to as the braiding function, characterises the kinetic mixing between the scalar and metric degrees of freedom and it is defined as \cite{Bellini2014}
\begin{equation}
\alpha_{\rm B}= \frac{\dot{\phi}XG_{3X}}{HM_{\rm pl}^2}\,.
\end{equation}
For the GCCG model, it is given by
\begin{equation}
\alpha_{\rm B}=
-sq\,\Omega_{\phi}^{0}
\left(\frac{H_0}{H}\right)^{s+2}.\label{eq:alphaB}
\end{equation}

The equation of motion for the scalar-field perturbation is 
\begin{equation}\label{eq:scalar_eom}
\partial^2\delta\phi
+
\frac{\lambda^2}{\dot{\phi}a^2}
\left[
\left(\partial_i\partial_j\delta\phi\right)^2-
\left(\partial^2\delta\phi\right)^2
\right]
=
-\frac{\lambda^2\dot{\phi}}{2M_{\rm pl}^2}
a^2\rho_{\rm m}\delta_{\rm m},
\end{equation}
where
\begin{equation}    
\lambda^2
=
-\frac{2\alpha_{\rm B}}{H\alpha c_s^2},
\end{equation}
and
\begin{equation}
\alpha c_s^2
=
2\left[
(1+\alpha_{\rm B})
\left(
-\frac{\dot H}{H^2}-\alpha_{\rm B}
\right)
-\frac{\dot{\alpha}_{\rm B}}{H}
\right]
-3\Omega_{\rm m}.   
\end{equation}
Here, $\alpha$ is the coefficient of the kinetic term in the quadratic action for scalar perturbations, while $c_s^2$ is the squared propagation speed of the scalar degree of freedom. The absence of ghost and gradient instabilities requires $
\alpha>0$, and $c_s^2>0$.
For the GCCG model considered here, the two gravitational potentials coincide within the QSA, so that $\Psi=\Phi$.

Following Ref.~\cite{Barreira:2013eea}, Eq.~\eqref{eq:scalar_eom} can be recast as
\begin{equation}\label{eq:scalar_eom2}
\partial^2\delta\phi
+
\frac{\left(\partial^2\delta\phi\right)^2
-
\left(\partial_i\partial_j\delta\phi\right)^2}{3\beta_1a^2\mathcal{M}^3}
=
\frac{8\pi G M_{\rm pl}}{3\beta_2}
a^2\rho_{\rm m}\delta_{\rm m},
\end{equation}
where $
\mathcal{M}^3=M_{\rm pl}H_0^2$,
and the two time-dependent functions $\beta_1$ and $\beta_2$ are defined as
\begin{eqnarray}
\beta_1
&=&
-\frac{\dot{\phi}}
{3\mathcal{M}^3\lambda^2},
\\
\beta_2
&=&
\frac{2\mathcal{M}^3M_{\rm pl}}
{\dot{\phi}^{2}}\beta_1.
\end{eqnarray}

We subsequently redefine the scalar-field perturbation according to
\begin{equation}
\delta\phi\rightarrow\frac{\beta}{\beta_2}\delta\phi,
\end{equation}
where $\beta$ is introduced for convenience. The scalar-field and Poisson equations then become
\begin{eqnarray}
\partial^2\delta\phi
+
\frac{\left(\partial^2\delta\phi\right)^2
-
\left(\partial_i\partial_j\delta\phi\right)^2}
{3(\beta_1\beta_2/\beta)a^2\mathcal{M}^3}
=
\frac{8\pi G M_{\rm pl}}{3\beta}
a^2\delta\rho_{\rm m},
\label{eq:scalar_eom_redefined}
\\
\partial^2\Psi
=
4\pi G a^2\rho_{\rm m}\delta_{\rm m}
+
\alpha_{\rm B}\frac{H}{\dot{\phi}}
\frac{\beta}{\beta_2}
\partial^2\delta\phi.
\label{eq:poisson_redefined}
\end{eqnarray}

The second term on the right-hand side of Eq.~\eqref{eq:poisson_redefined} represents the modification of the gravitational force mediated by the scalar degree of freedom, commonly referred to as the fifth force. For a spherically symmetric configuration, in which both the gravitational and scalar fields depend only on the radial coordinate (r), its radial component is
\begin{equation}\label{eq:5thforce}
F_{\rm 5th}=
\alpha_{\rm B}\frac{H}{\dot{\phi}}
\frac{\beta}{\beta_2}
\frac{\partial\delta\phi}{\partial r}.
\end{equation}

Equation~\eqref{eq:scalar_eom_redefined} is a nonlinear elliptic equation in which the quadratic derivative term encodes the Vainshtein screening mechanism. In low-density regions, this contribution is subdominant and the scalar field mediates an additional, unscreened gravitational interaction. In highly nonlinear environments, the derivative self-interactions suppress the scalar-field gradient and hence the fifth force in Eq.~\eqref{eq:5thforce}. The strength of this force is controlled by the background-dependent functions $\alpha_{\rm B}$. Solving the full nonlinear equation in the N-body code therefore allows the environmental dependence of screening to be captured self-consistently.

Finally, Eqs.~\eqref{eq:scalar_eom_redefined}, \eqref{eq:poisson_redefined}, and \eqref{eq:5thforce} are the equations implemented in our N-body code. We emphasise that their structure is general for shift-symmetric cubic Galileon theories. The code can therefore be readily adapted to models described by a different cubic Lagrangian by modifying the corresponding background-dependent functions. Within the formulation adopted here, the GCCG-specific model dependence enters through the braiding function $\alpha_{\rm B}$, whose explicit expression is given in Eq.~\eqref{eq:alphaB}.

\section{N-body simulations set-up}\label{sec:Nbody}

We implement the GCCG model in the \texttt{ECOSMOG} simulation code \cite{Li:2011vk,Li:2013nua}, which is an adaptive mesh refinement (AMR) code built on \texttt{Ramses} \cite{Teyssier:2001cp} and extended to model modified gravity scenarios. It employs multigrid relaxation methods to solve the non-linear Klein--Gordon equations governing the additional scalar fields that arise in MG theories. For details on the code algorithm, we refer the reader to Refs. \cite{Li:2011vk,Barreira:2013eea}. \texttt{ECOSMOG} has been used to simulate several models, such as Hu-Sawicki $f(R)$ gravity \citep{Li:2011vk}, nDGP \cite{Li:2013nua}, the symmetron mechanism \citep{Davis:2011pj,Brax:2013mua}, dilaton theories \citep{Brax:2011ja}, and galileon gravity, including cubic, quartic, and cubic vector formulations \citep{Barreira:2013eea,Becker:2020azq}. 

Equations~\eqref{eq:scalar_eom_redefined}, \eqref{eq:poisson_redefined}, and \eqref{eq:5thforce} are implemented in the N-body code. The background evolution, namely $H$,  is precomputed numerically and stored in a table, from which the code obtains the required background quantities through interpolation. Since the nonlinear equations retain the same structure for shift-symmetric cubic Galileon theories, this tabulated approach allows the implementation to be adapted to a different model by replacing the corresponding background functions and the form of $\alpha_{\rm B}$. 

 For the GCCG simulations, we adopt $q=1.06$ and $s=0.65$, together with $\Omega_{\rm b}^{0}h^2=0.0225$, $\Omega_{\rm c}^{0}h^2=0.117$, $h=H_0/(100\, \mbox{km}\,\mbox{s}^{-1}\mbox{Mpc}^{-1})=0.7381$, $n_s=0.973$, $\tau=0.05$, and $\ln(10^{10}A_s)=3.058$ at the pivot scale $k_{\rm piv}=0.02\,{\rm Mpc}^{-1}$. These values correspond to the best-fit GCCG cosmology~\cite{Frusciante:2019puu}. 

In addition to the GCCG simulations, we consider a corresponding QCDM model, in which the gravitational force is computed according to GR while the background expansion history is fixed to that of GCCG. Comparison with this baseline isolates the impact of the MG interaction on the growth of matter perturbations from that induced solely by the modified expansion history. We also perform $\Lambda$CDM simulations with the same set-up as the GCCG simulations.

 We perform dark-matter-only simulations in cubic boxes of side length $L=1024,h^{-1}{\rm Mpc}$. The simulations contain $N_{\rm p}=1024^3$ particles, while the domain grid consists of $1024^3$ cells. The refinement threshold is set to $N_{{\rm p},{\rm th}}=10$.

The initial conditions were generated at $z_{\rm ini}=49$ using a
modified version of \texttt{2LPTic}~\cite{Crocce:2006ve}. In its
standard implementation, \texttt{2LPTic} receives the linear matter
power spectrum specified at $z=0$ and rescales it to the initial
redshift using the assumed linear growth history. In the version
adopted here, this back-scaling step is bypassed: the code directly
reads an externally generated linear matter power spectrum evaluated
at $z_{\rm ini}$. The initial density and displacement-field
amplitudes are therefore determined directly from
$P(k,z_{\rm ini})$, computed with \texttt{EFTCAMB} \cite{Frusciante:2019puu}.

For the GCCG simulation, we used the corresponding model-specific
linear matter power spectrum, whereas the QCDM and $\Lambda$CDM
simulations were initialised using the same $\Lambda$CDM linear matter
power spectrum evaluated at $z_{\rm ini}$. We performed one realisation
for each cosmology. The three simulations were generated using the same
Gaussian random phases and therefore form a matched set, reducing sample
variance in comparisons between the different cosmologies. The
displacement-field amplitudes were determined by the initial matter
power spectrum adopted for each model, while the peculiar velocities
were assigned at $z_{\rm ini}$ using the standard \texttt{2LPTic}
prescription.

At $z_{\rm ini}=49$, the GCCG and $\Lambda$CDM linear matter power spectra are indistinguishable, within numerical precision, over the
range of wavenumbers relevant to the simulations. Consequently, the
GCCG and QCDM simulations start from effectively identical initial density, displacement, and velocity fields. Since they also share the same background expansion history, their comparison isolates, to
numerical precision, the subsequent effect of the scalar-mediated interaction.

The GCCG implementation used in this work has been independently
cross-validated against the \texttt{EFT-RAMSES} implementation~\cite{Woodcock:2026lpv}. While \texttt{EFT-RAMSES} maps the GCCG
model onto a general EFT-based master equation, the implementation
adopted here employs an alternative scalar-field redefinition and
computes the background-dependent coefficients from numerically
pre-calculated tables. Using matched initial conditions and identical
simulation specifications, Ref.~\cite{Woodcock:2026lpv} compared the
absolute GCCG and QCDM matter power spectra, as well as the ratio
$P_{\rm GCCG}/P_{\rm QCDM}$, at $z=0$, finding excellent agreement
between the two implementations over the range of scales considered.
The power-spectrum data used in that comparison correspond to a
preliminary version of the simulations analysed in detail in the
present work. This comparison therefore provides a non-trivial validation of the
nonlinear scalar-field solver, of the background-dependent
coefficients, and of the normalization of the fifth-force contribution
adopted in our simulations.

\begin{figure*}[ht!]
    \centering
\includegraphics[width=1.\textwidth]{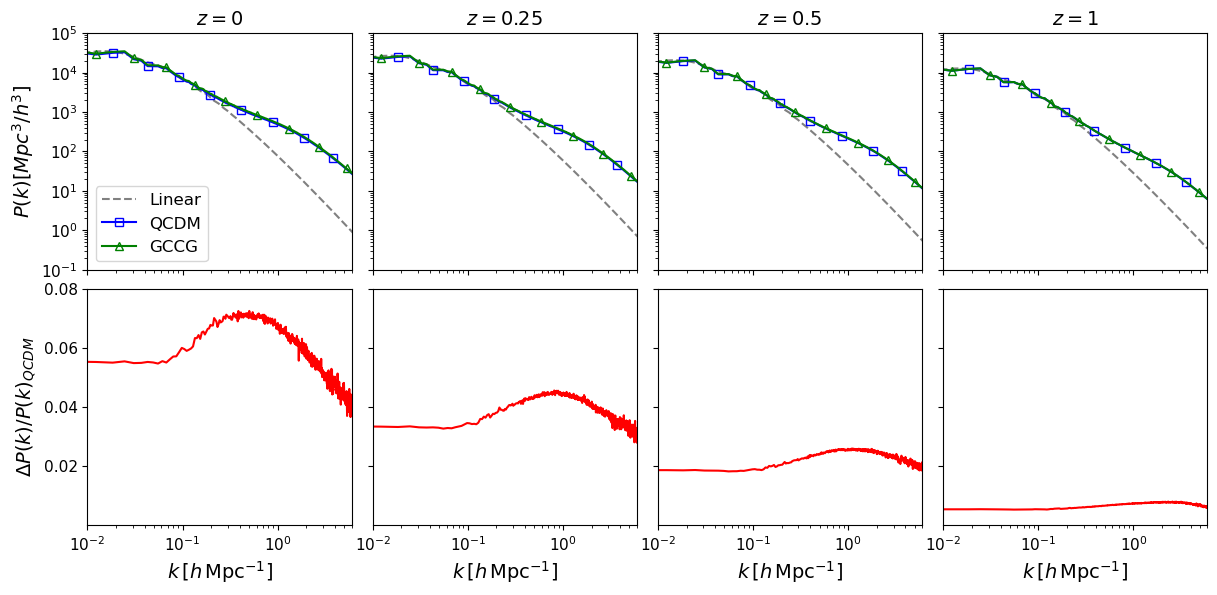}
   \caption{
Matter power spectrum in the GCCG and QCDM cosmologies at
$z=0$, $0.25$, $0.5$, and $1$, from left to right. The upper panels show the
measurements from the $N$-body simulations, together with the linear GCCG
prediction obtained with \texttt{EFTCAMB}. The lower panels display the
fractional difference
$\left[P_{\rm GCCG}(k,z)-P_{\rm QCDM}(k,z)\right]/P_{\rm QCDM}(k,z)$. 
}
    \label{fig:pk}
\end{figure*}

\section{Results}\label{sec:results}

We now present the main results of our $N$-body simulations. We begin by
examining the GCCG nonlinear matter power spectrum and quantifying its departures
from the corresponding QCDM and $\Lambda$CDM reference cosmologies as a
function of scale and redshift. We then assess the accuracy of the predictions
obtained with the halo-model reaction formalism of \texttt{ReACT}, using the GCCG implementation introduced in
Ref.~\cite{Atayde:2024tnr}, through a direct comparison with the \texttt{ECOSMOG} simulation results. Finally, we investigate the impact of the GCCG dynamics on the abundance of collapsed structures through the cumulative
halo mass function.

For compatibility with the
following analysis pipeline, the \texttt{ECOSMOG} outputs, written in the
native \texttt{Ramses}~\cite{Teyssier:2001cp} format, were converted to the \texttt{Gadget} ~\cite{Springel:2005mi} format
using \texttt{Ramses2Gadget}. The matter power spectra were measured with
\texttt{Powmes}~\cite{2011ascl.soft10017C}, while the halo catalogue was
constructed using \texttt{ROCKSTAR}~\cite{2013ApJ...762..109B}.
Visualisations of the matter density and velocity fields were produced with
the \texttt{Pynbody} Python package~\cite{2013ascl.soft05002P}.

\subsection{Nonlinear matter power spectrum}
\label{sec:pk_results}

The upper panels of Fig.~\ref{fig:pk} show the matter power spectra measured
from the GCCG and QCDM simulations at $z=0$, $0.25$, $0.5$, and $1$,
together with the linear GCCG prediction obtained with
\texttt{EFTCAMB} \cite{Frusciante:2019puu}. The QCDM cosmology has the same background expansion
history and cosmological parameters as GCCG, while the evolution of matter
perturbations is governed by GR. The comparison between the two simulations
therefore isolates the effect of the scalar-mediated interaction on the
growth of structures.

At the smallest wavenumbers probed by the simulations,
$k\simeq 0.01\,h\,{\rm Mpc}^{-1}$, the measured spectra are consistent with
the linear-theory prediction. Towards larger wavenumbers, nonlinear evolution
progressively enhances the power relative to linear theory and generates a
nontrivial scale dependence in the difference between the GCCG and QCDM
predictions. To quantify this effect, the lower panels of Fig.~\ref{fig:pk} show the
fractional difference
\begin{equation}
    \frac{\Delta P}{P_{\rm QCDM}}
    \equiv
    \frac{P_{\rm GCCG}(k,z)-P_{\rm QCDM}(k,z)}
         {P_{\rm QCDM}(k,z)} .
    \label{eq:pk_fractional_difference}
\end{equation}
The GCCG model enhances the matter power spectrum at all the redshifts
considered, with the amplitude of the modification increasing towards late
times. On the largest scales shown, the enhancement is approximately
$5.5\%$ at $z=0$, $3.5\%$ at $z=0.25$, $2\%$ at $z=0.5$, and below
$1\%$ at $z=1$. This redshift dependence reflects the late-time growth of
the scalar-mediated force, whose impact on matter clustering becomes progressively more significant as the GCCG dark-energy component becomes dynamically important.

The scale dependence becomes more pronounced once nonlinear evolution sets
in. Starting from its large-scale value, the fractional enhancement initially
increases with wavenumber and reaches a broad maximum in the transition
between the quasilinear and nonlinear regimes. At $z=0$, the enhancement
peaks at approximately $7\%$, while the corresponding maxima are about
$4.5\%$, $2.5\%$, and $1\%$ at $z=0.25$, $0.5$, and $1$, respectively.
The position of the maximum also shifts towards smaller wavenumbers at later
times, consistently with nonlinear evolution affecting progressively larger
physical scales.

Beyond the maximum, the enhancement decreases towards larger wavenumbers.
This turnover is the expected imprint of Vainshtein screening. As increasingly
nonlinear and high-density regions contribute to the power spectrum, the
derivative self-interactions of the scalar field suppress its spatial
gradients and reduce the contribution of the fifth force to matter
clustering. The shape of the GCCG-to-QCDM ratio therefore reflects the
competition between the nonlinear amplification of the modified gravitational
interaction and its suppression by the screening mechanism.

At the largest wavenumbers probed by the simulations, the GCCG spectrum
nevertheless remains above its QCDM counterpart. The residual enhancement is
of order $4\%$ at $z=0$, $3\%$ at $z=0.25$, and $2\%$ at $z=0.5$, while
it remains below the percent level at $z=1$. The range of scales covered by
the simulations is therefore sufficient to identify the onset and increasing
efficiency of Vainshtein screening, but does not establish whether the
GCCG-to-QCDM ratio asymptotically approaches unity on still smaller scales.

\subsection{Comparison with the halo-model reaction}
\label{sec:react_results}

The dedicated GCCG simulations presented here enable a direct assessment of
the halo-model reaction implementation introduced in
Ref.~\cite{Atayde:2024tnr}. In that work, the implementation was tested against
existing $N$-body simulations only in the Cubic Galileon limit,
$s=2$ and $q=1/2$, because simulations of the general GCCG model were not
available. The present analysis therefore provides the first validation of
the reaction prediction for a GCCG cosmology.

We characterize the relative impact of a
cosmological model on nonlinear clustering through the power-spectrum boost.
For two cosmologies $X$ and $Y$, this is defined as
\begin{equation}
    B_{X/Y}(k,z)
    \equiv
    \frac{P^{X}_{\rm NL}(k,z)}
         {P^{Y}_{\rm NL}(k,z)} .
    \label{eq:boost_definition}
\end{equation}
The boost therefore measures the scale- and redshift-dependent change in the
nonlinear matter power spectrum of cosmology $X$ relative to the reference
cosmology $Y$. A value $B_{X/Y}=1$ indicates identical clustering in the two
models, whereas $B_{X/Y}>1$ and $B_{X/Y}<1$ correspond, respectively, to an
enhancement and a suppression of power in $X$ relative to $Y$.

We assess the accuracy of the halo-model reaction by comparing the boost
predicted by \texttt{ReACT} with that measured from the simulations,
\begin{equation}
    \mathcal{Q}_{X/Y}(k,z)
    \equiv
    \frac{B_{X/Y}^{\textsc{ReACT}}(k,z)}
         {B_{X/Y}^{\textsc{ECOSMOG}}(k,z)} .
    \label{eq:react_ecosmog_ratio}
\end{equation}
Accordingly, $\mathcal{Q}_{X/Y}=1$ denotes perfect agreement, while departures
from unity quantify the fractional modelling error in the predicted boost.

\begin{figure*}[ht!]
    \centering
\includegraphics[width=1.\textwidth]{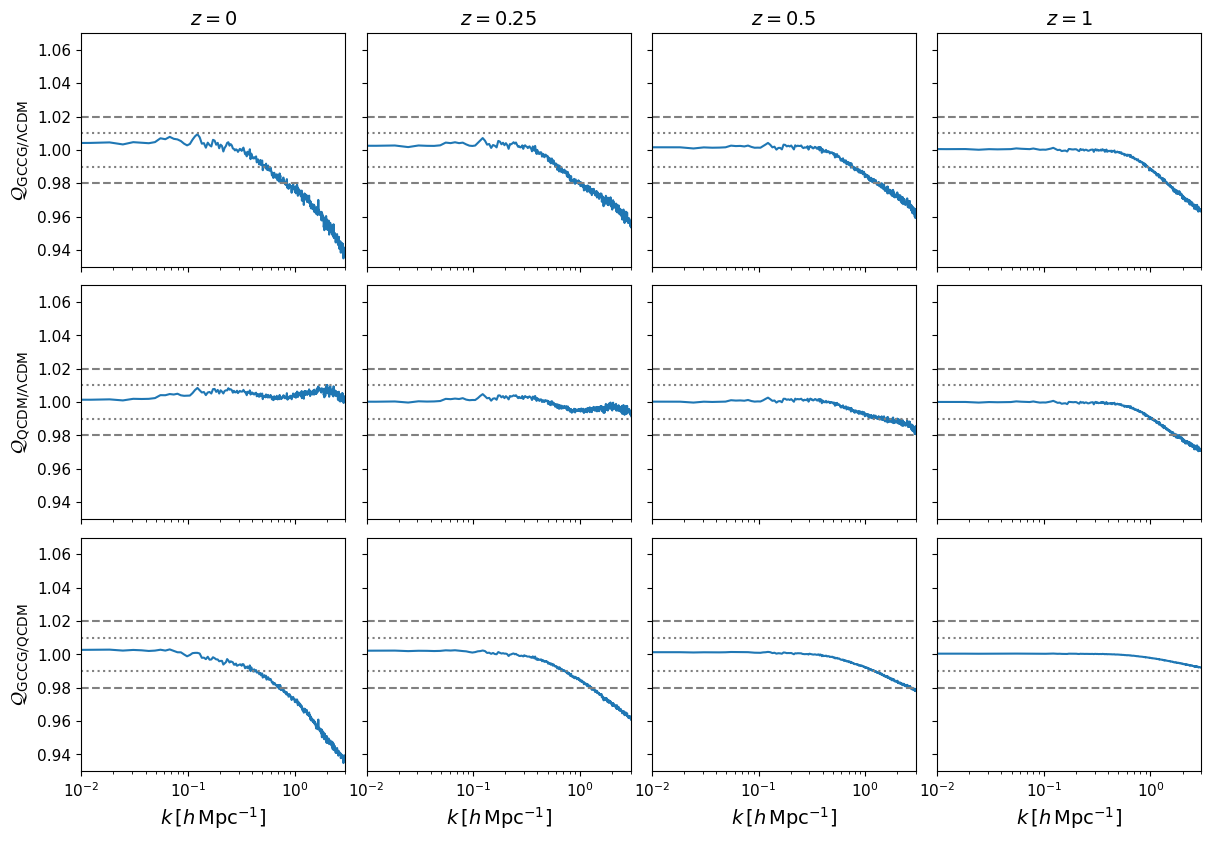}
    \caption{Comparison between the nonlinear power-spectrum boosts
predicted by \texttt{ReACT} and those measured from the
\texttt{ECOSMOG} simulations. The panels show the ratio
$\mathcal{Q}_{X/Y}=B^{\texttt{ReACT}}_{X/Y}/B^{\texttt{ECOSMOG}}_{X/Y}$ at
$z=0$, $0.25$, $0.5$, and $1$, from left to right. From top to bottom,
the rows correspond to the GCCG-to-$\Lambda$CDM,
QCDM-to-$\Lambda$CDM, and GCCG-to-QCDM boosts, respectively. A value
$\mathcal{Q}_{X/Y}=1$ indicates perfect agreement between the reaction
prediction and the simulation result. The dotted and dashed horizontal
lines mark deviations of $1\%$ and $2\%$ from unity, respectively.}
    \label{fig:react_comparison}
\end{figure*}

Figure~\ref{fig:react_comparison} shows this quantity for the
GCCG-to-$\Lambda$CDM boost in the upper panels, the
QCDM-to-$\Lambda$CDM boost in the middle panels, and the GCCG-to-QCDM
boost in the lower panels, at $z=0$, $0.25$, $0.5$, and $1$. The dotted
and dashed horizontal lines indicate deviations of $1\%$ and $2\%$ from
the simulation results, respectively.

The three sets of ratios separate the different contributions entering the
full GCCG prediction. The GCCG-to-$\Lambda$CDM boost includes both the
effect of the GCCG expansion history and that of the scalar-mediated
interaction. The QCDM-to-$\Lambda$CDM ratio instead probes the nonlinear
effect of the modified background evolution in the absence of the fifth
force. Since GCCG and QCDM share the same background expansion history, the
GCCG-to-QCDM boost isolates the modification associated with the scalar
degree of freedom.

On the largest scales, the reaction predictions approach the simulation
measurements for all three boosts. The agreement remains at the percent level
through the linear and mildly nonlinear regimes, whereas systematic
differences emerge towards larger wavenumbers. For the full
GCCG-to-$\Lambda$CDM boost, \texttt{ReACT} increasingly underpredicts the
simulation result as nonlinear evolution becomes important. The discrepancy
is largest at $z=0$, where it reaches approximately $5$--$6\%$ at the
highest wavenumbers shown. It decreases to about $3$--$4\%$ at $z=0.25$
and remains at the few-percent level at $z=0.5$ and $z=1$.

The QCDM-to-$\Lambda$CDM comparison displays a substantially weaker
departure from unity at low redshift. At $z=0$ and $z=0.25$, the reaction
prediction remains close to the simulation result over most of the scale
range, with deviations at the percent level. The discrepancy becomes more
pronounced at higher redshift and at the largest wavenumbers, reaching a few
percent at $z=1$. This shows that the nonlinear modelling of the background
contribution alone cannot account for the disagreement observed in the full
GCCG boost at low redshift.

This is confirmed by the GCCG-to-QCDM ratio in the lower panels. At $z=0$,
the reaction and simulation results agree on large scales, but the ratio
decreases steadily once nonlinear scales are reached, approaching
approximately $0.94$ at the largest wavenumbers considered. The discrepancy
decreases with increasing redshift, amounting to approximately $3$--$4\%$
at $z=0.25$, around $2\%$ at $z=0.5$, and less than $1\%$ at $z=1$.
Thus, at low redshift, most of the difference in the full
GCCG-to-$\Lambda$CDM boost originates from an underestimation of the
nonlinear modified-gravity contribution. At $z=1$, by contrast, the
GCCG-to-QCDM boost is accurately reproduced, and the residual discrepancy in
the full boost is mainly inherited from the QCDM-to-$\Lambda$CDM prediction.

More quantitatively, at $z=0$ the \texttt{ReACT} prediction for the GCCG-to-$\Lambda$CDM boost remains within $1\%$ of the simulation result up to $k \simeq 0.50\,h\,{\rm Mpc}^{-1}$ and within $2\%$ up to $k \simeq 0.81\,h\,{\rm Mpc}^{-1}$. For the GCCG-to-QCDM boost, the corresponding scales are $k \simeq 0.39\,h\,{\rm Mpc}^{-1}$ and $k \simeq 0.71\,h\,{\rm Mpc}^{-1}$, respectively.

Overall, the halo-model reaction reproduces the scale and redshift dependence of the simulated GCCG boosts and provides a percent-level description on linear and mildly nonlinear scales. Its accuracy deteriorates to several percent at low redshift in the deeply nonlinear regime, where \texttt{ReACT} underestimates the residual enhancement produced by the scalar-mediated interaction. The comparison does not, by itself, identify which individual ingredient of the reaction calculation is responsible for this discrepancy. Nevertheless, it provides the first simulation-based determination of the accuracy and range of validity of the halo-model reaction for a genuine GCCG cosmology.

\begin{figure*}[ht!]
    \centering
\includegraphics[width=1.\textwidth]{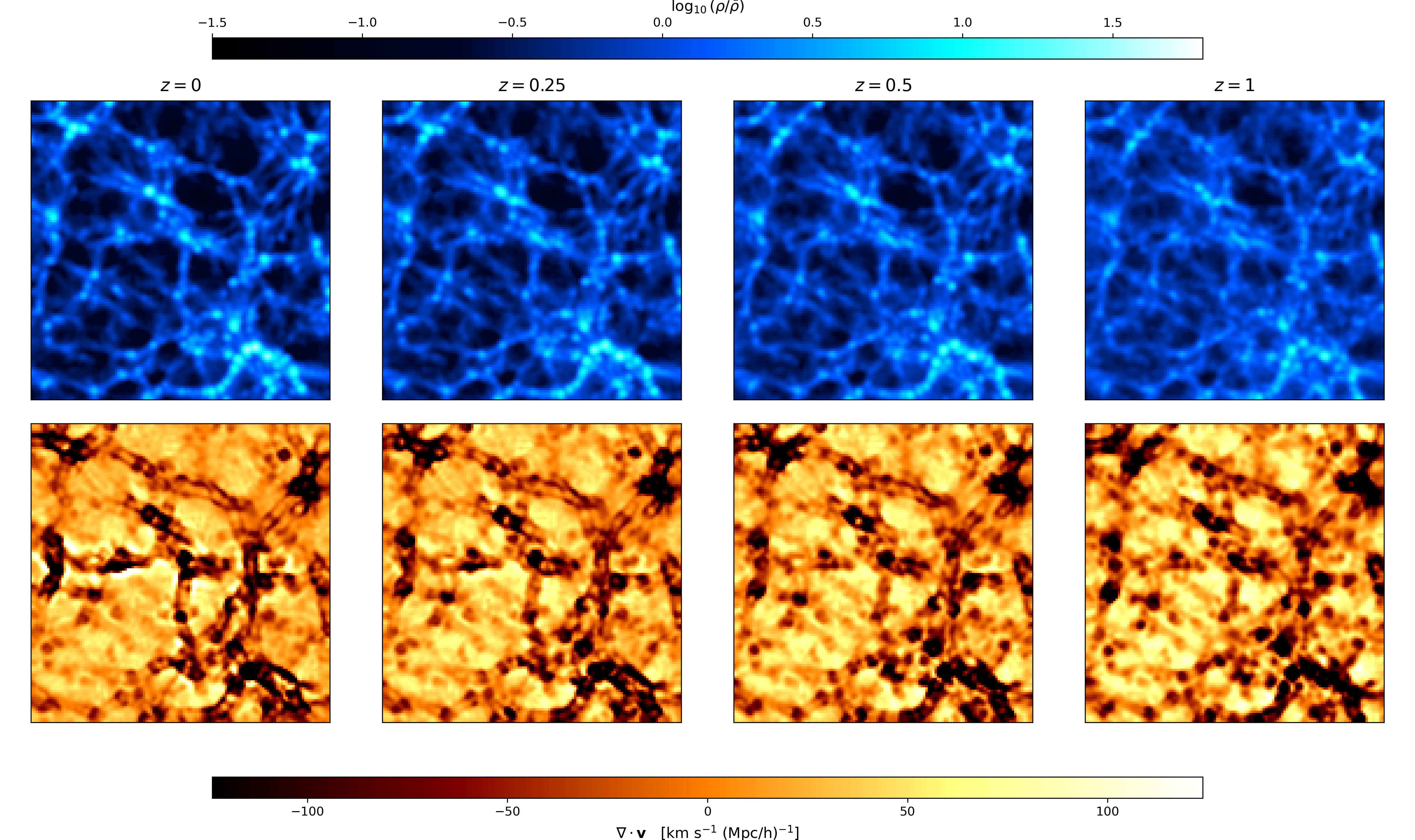}
   \caption{
Evolution of the matter density and peculiar-velocity fields in the GCCG
simulation. The maps show the same
$128\,h^{-1}{\rm Mpc}\times128\,h^{-1}{\rm Mpc}\times 12\,h^{-1}{\rm Mpc}$ slice centred on the
simulation box at  $z=0$, $0.25$, $0.5$, and $1$. The upper
panels display the logarithmic matter density,
$\log_{10}(\rho_{\rm m}/\bar{\rho}_{\rm m})$, while the lower panels show the
divergence of the peculiar-velocity field. Negative velocity divergence
traces convergent flows towards filaments and overdense nodes, whereas
positive values predominantly trace outflows from underdense regions.
}
\label{fig:density_velocity}
\end{figure*}

\subsection{Evolution of the density and velocity fields}
\label{sec:fields}

To complement the statistical characterisation of nonlinear structure
formation, Fig.~\ref{fig:density_velocity} shows the spatial evolution of the
matter density and velocity fields in the GCCG simulation. The maps correspond
to the same $128\,h^{-1}{\rm Mpc} \times 128\,h^{-1}{\rm Mpc} \times 12\,h^{-1}{\rm Mpc}$ slice centred on the simulation box and are
shown at 
$0$, $0.25$, $0.5$, and $z=1$, respectively. The upper panels display
$\log_{10}(\rho_{\rm m}/\bar{\rho}_{\rm m})$, while the lower panels show the
divergence of the peculiar-velocity field.

The cosmic-web morphology is already clearly established at $z=1$, with
matter distributed in a network of sheets, filaments, and compact nodes
surrounding underdense regions. As the system evolves towards $z=0$, the
density contrast increases: filaments become more sharply defined, dense
knots grow more prominent, and matter is progressively depleted from the
intervening low-density regions. These maps provide a direct visual
illustration of the nonlinear growth responsible for the scale-dependent
enhancement of the matter power spectrum discussed in
Sec.~\ref{sec:pk_results}.

The velocity-divergence field exhibits a closely related spatial pattern.
With the sign convention adopted here, negative values identify convergent
flows and are preferentially found around filaments and high-density nodes,
where matter is undergoing gravitational infall. Positive velocity divergence
is instead predominantly associated with underdense regions, reflecting the
outflow of matter from voids towards the surrounding structures. The
amplitude and spatial coherence of these flows become increasingly pronounced
at late times, in parallel with the growth of the density contrast.

The correspondence between the density and velocity fields illustrates the
dynamical assembly of the cosmic web in the GCCG cosmology. 

\subsection{Halo abundance}
\label{sec:hmf_results}

\begin{figure*}[ht!]
    \centering
\includegraphics[width=1.\textwidth]{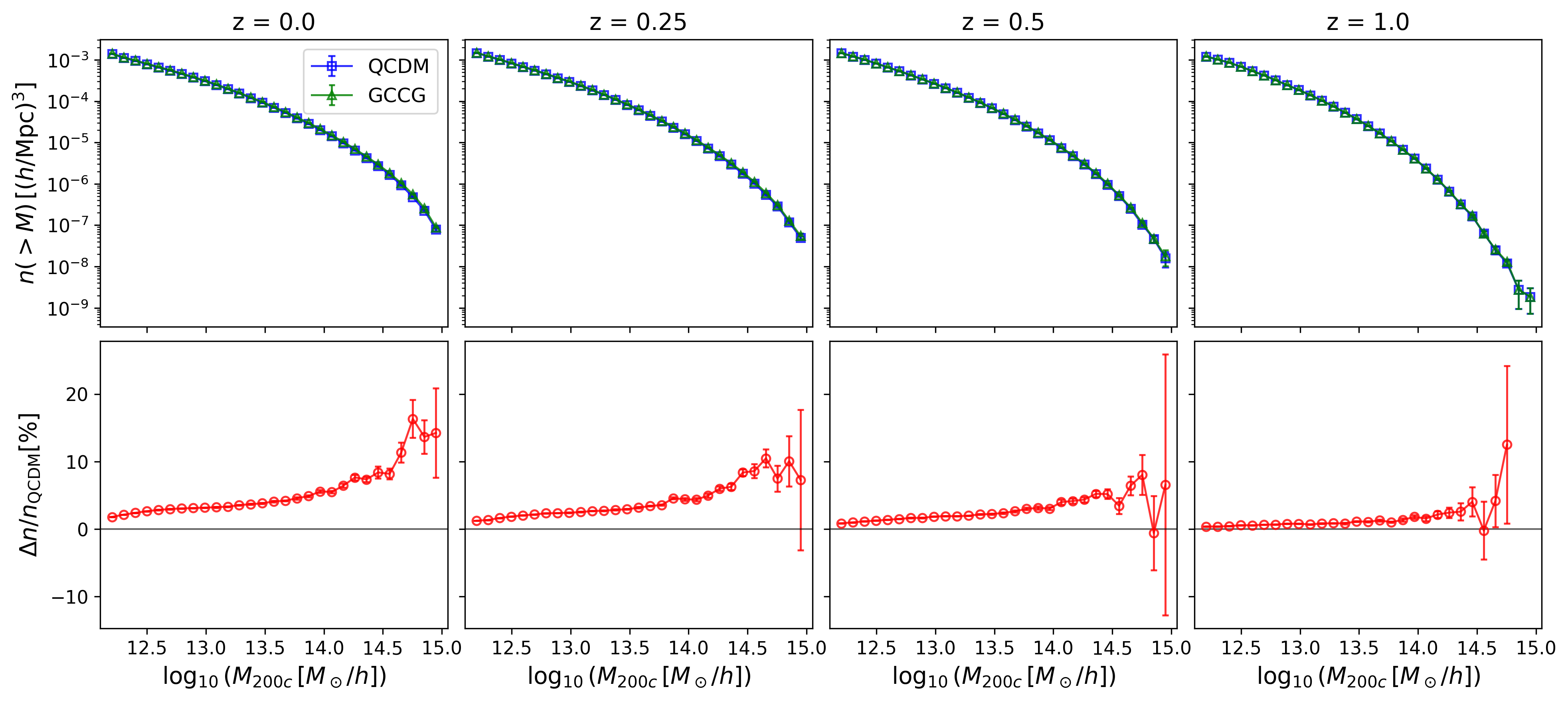}
    \caption{Cumulative halo mass function in the GCCG and QCDM simulations at
$z=0$, $0.25$, $0.5$, and $1$, from left to right. The upper panels show
the  number density of haloes with mass larger than
$M_{200{\rm c}}$, while the lower panels display the relative difference.}
    \label{fig:hmf}
\end{figure*}

The impact of the GCCG interaction on collapsed objects is shown in
Fig.~\ref{fig:hmf}. The upper panels present the cumulative halo mass function,
$n(>M_{200{\rm c}})$, for the GCCG and QCDM simulations. The lower panels show
the relative difference
\begin{equation}
    \frac{\Delta n}{n_{\rm QCDM}}
    \equiv
    \frac{n_{\rm GCCG}(>M_{200{\rm c}})
          -n_{\rm QCDM}(>M_{200{\rm c}})}
         {n_{\rm QCDM}(>M_{200{\rm c}})},
    \label{eq:hmf_fractional_difference}
\end{equation}
where the halo mass $M_{200{\rm c}}$ is defined as the mass enclosed within the
radius $R_{200{\rm c}}$ for which the mean interior density is $200$ times
the critical density at redshift $z$,
\begin{equation}
M_{200{\rm c}}
=
\frac{4\pi}{3}\,200\,\rho_{\rm c}(z)\,R_{200{\rm c}}^{3},
\end{equation}
and
\begin{equation}
\rho_{\rm c}(z)=\frac{3H^{2}(z)}{8\pi G}.
\end{equation}

The results are shown as the mean over eight spatial sub-volumes of the simulation box, with error bars representing the sub-volume sample scatter.

The GCCG model predicts a larger cumulative halo abundance than QCDM over
almost the entire resolved mass range. The difference is small for
low-mass haloes but increases systematically toward the exponential tail of
the mass function. It also becomes larger at later times, consistently with
the redshift evolution of the matter power spectrum.

At $z=0$, the enhancement is at the level of a few percent for
$M_{200{\rm c}}\lesssim 10^{13}\,h^{-1}M_\odot$, increases above
$M_{200{\rm c}}\sim10^{14}\,h^{-1}M_\odot$, and reaches approximately
$10$--$15\%$ in the highest sufficiently populated mass bins. At
$z=0.25$ and $z=0.5$, the enhancement in the massive-halo regime is smaller,
typically of order $5$--$10\%$. At $z=1$, the two mass functions are much
closer, with differences of only a few percent except in the sparsely
populated high-mass tail.

The increase in the relative difference with halo mass is expected from the
exponential sensitivity of rare-object abundances to the amplitude and growth
rate of matter fluctuations. A modest enhancement of the matter power
spectrum can therefore produce a substantially larger fractional change in
the number of massive haloes. The positive GCCG correction is consistent with
the additional attractive interaction inferred from the power spectrum
results: matter collapses more efficiently than in the QCDM cosmology, leading
to an increased number of objects above a fixed mass threshold.

The final mass bins at each redshift contain only a small number of objects
and consequently display large uncertainties. In particular, isolated
fluctuations in these bins should not be assigned physical significance. The
robust result is the coherent enhancement extending over several adjacent
mass bins and its systematic increase toward low redshift and high halo mass.

Taken together, the power spectrum and halo-abundance measurements show that
the GCCG model produces a late-time enhancement of structure formation whose
scale dependence is regulated, but not completely erased, by Vainshtein
screening. 

\section{Conclusion}\label{sec:conclusion}

In this work, we have presented the first dedicated $N$-body simulations of
large-scale structure formation in the Generalized Cubic Covariant Galileon
model. We implemented the nonlinear scalar-field equation in the
adaptive-mesh-refinement code \texttt{ECOSMOG}, allowing the matter
distribution and the scalar degree of freedom to be evolved self-consistently
in the presence of Vainshtein screening. In addition to the full GCCG
simulations, we considered a QCDM cosmology with the same background expansion
history but with matter perturbations governed by GR, as well as a reference
$\Lambda$CDM cosmology. These complementary simulations make it possible to
separate the effects of the modified background evolution from those produced
by the scalar-mediated interaction.

For the GCCG parameter values considered here, the scalar interaction enhances
the nonlinear matter power spectrum relative to QCDM at all the redshifts
analysed. The modification becomes progressively larger towards late times,
reaching approximately $5.5\%$ on the largest scales probed and a maximum of
about $7\%$ in the transition between the quasilinear and nonlinear regimes
at $z=0$. Beyond this maximum, the enhancement decreases towards larger
wavenumbers. This turnover is consistent with the increasing efficiency of
Vainshtein screening in nonlinear, high-density environments. Nevertheless,
the GCCG power spectrum remains above the QCDM result throughout the range of
scales resolved by the simulations, with a residual enhancement of
approximately $4\%$ at the largest wavenumbers shown at $z=0$.

The simulations also enabled the first direct assessment of the GCCG
implementation in the halo-model reaction code \texttt{ReACT} away from the
Cubic Galileon limit. We compared the nonlinear power-spectrum boosts for
GCCG relative to both QCDM and $\Lambda$CDM, as well as the corresponding
QCDM-to-$\Lambda$CDM boost. The reaction formalism reproduces the large-scale
normalisation and the main scale and redshift dependence measured in the
simulations. Its predictions remain accurate at approximately the percent
level over the linear and mildly nonlinear regimes. At low redshift, however,
\texttt{ReACT} increasingly underestimates the simulated GCCG enhancement
towards deeply nonlinear scales. For the GCCG-to-QCDM boost, the discrepancy
reaches approximately $5$--$6\%$ at $z=0$, decreases to about $3$--$4\%$ at
$z=0.25$ and $2\%$ at $z=0.5$, and remains below the percent level at
$z=1$. The comparison indicates that the dominant low-redshift discrepancy
arises from the modelling of the nonlinear modified-gravity contribution,
rather than from the GCCG background evolution alone.

The enhanced growth of structure is also reflected in the abundance of
collapsed objects. The cumulative halo mass function is systematically larger
in GCCG than in QCDM over most of the resolved mass range. The relative
difference increases towards lower redshift and higher halo mass, reaching
approximately $10$--$15\%$ in the highest sufficiently populated mass bins at
$z=0$. This behaviour is consistent with the enhanced matter power spectrum:
even a moderate change in the amplitude of matter fluctuations can lead to a
larger fractional variation in the exponentially suppressed high-mass tail of
the halo population. The density and velocity-divergence maps provide a
complementary qualitative illustration of the assembly of the cosmic web,
showing the progressive sharpening of filaments and overdense nodes and the
associated convergent and divergent peculiar-velocity flows.

The present analysis is restricted to a single observationally motivated
GCCG parameter point and to dark-matter-only simulations. The numerical
results therefore constitute an initial nonlinear calibration rather than a
complete exploration of the GCCG parameter space. A broader simulation suite
will be required to determine how the nonlinear signatures and the efficiency
of Vainshtein screening vary across the allowed model space. Such simulations
would also permit a systematic recalibration of the reaction ingredients and
the construction of fast emulators suitable for likelihood analyses.

Overall, our results demonstrate that the GCCG interaction leaves measurable
signatures in nonlinear matter clustering and halo abundance, while
Vainshtein screening regulates these deviations on small scales without
completely erasing them over the range probed here. The simulations provide
the first numerical benchmark for nonlinear structure formation in GCCG and constitute a necessary step towards robust tests
of this class of MG models with forthcoming large-scale
structure observations from Euclid, Rubin/LSST, Roman, and SKAO.

\acknowledgments

This work used the DiRAC@Durham facility managed by the Institute for Computational Cosmology on behalf of the STFC DiRAC HPC Facility (https://www.dirac.ac.uk). The equipment was funded by BEIS capital funding via STFC capital grants ST/K00042X/1, ST/P002293/1, ST/R002371/1 and ST/S002502/1, Durham University and STFC operations grant ST/R000832/1. DiRAC is part of the National e-Infrastructure.

L.A. is supported by Fundação para a Ciência e a Tecnologia (FCT) through the research grants UIDB/04434/2020, UIDP/04434/2020 and  from the FCT PhD fellowship grant with ref. number 2022.11152.BD.
L.A. and N.F.  also acknowledge the FCT project with ref. number PTDC/FIS-AST/0054/2021 and  the COST Action CosmoVerse, CA21136, supported by COST (European Cooperation in Science and Technology).
 N.F. acknowledges the Istituto Nazionale di Fisica Nucleare (INFN) Sez. di Napoli, Iniziativa Specifica InDark. 
B.L.~is supported by the European Research Council (ERC) Advanced Grant ``UNCA'', under the UKRI's Frontiers Research Guarantee, Grant No.~EP/Z533877/1, and the UK STFC Consolidated Grant No.~ST/X001075/1.

\appendix

\bibliography{bib}

\end{document}